\documentstyle[preprint,aps]{revtex}
\begin{document}
\draft
\title{Cytoskeleton influence on
normal and tangent fluctuation modes in the red blood cells}
\author{S.B. Rochal$^{\dagger,\ddagger}$ and  V.L. Lorman$^\ddagger$}
\address{$^\ddagger$Laboratoire de Physique  Theorique  et Astroparticules, CNRS - Universite
Montpellier 2, Place Eugene Bataillon, 34095 Montpellier, France}
\address{$^\dagger$Physical Faculty, Rostov State University, 5 Zorge Street,
         344090 Rostov-on-Don, Russia}

\date{\today}
\maketitle

\begin{abstract}
We argue that the paradoxal softness of the red blood cells (RBC)
in fluctuation spectra experiments is apparent. We show that the
effective surface shear modulus $\mu_s$ of the RBC obtained from
fluctuation data and that measured in static deformation
experiments have the same order of magnitude. A simple
micromechanical model of the RBC developped for this purpose
accounts  for the influence of a finite-thickness cytoskeleton on
the fluctuations of the composite membrane-cytoskeleton system.
The spectrin network cytoskeleton with the bulk shear modulus
estimated as $\mu\approx105\div 165$ Pa contributes  to both normal
and tangent fluctuations of the system and confines the
fluctuations of the lipid membrane. The ratio of mean square
amplitudes of the RBC normal and tangent fluctuations $<X_n^2> /< X_t^2>$
calculated in the frame of the model is 2-3 orders of magnitude
smaller that it is in the free membrane with the same bending and
shear moduli.

\end{abstract}

\pacs{87.68.+z, 83.60.-a, 87.16.-b, 87.17.-d}

\renewcommand{\vec}[1]{\mbox{\boldmath$#1$}}
\newcommand{\tr}[1]{\mbox{\boldmath$#1$}}
\newcommand{\indE}{\mbox{\scriptsize$E$}}
\newcommand{\indEw}{E_{kl},\partial_kw_l}

The mechanical properties and fluctuation spectrum of the living cell depend
strongly on the coupling between its fluid phospholipid membrane (ME) and stiff
cytoskeleton (CS) composed of the cross-linked biopolimer networks [1-4]. In red
blood cells (RBC), investigated intensively by physicists since many years [5-7],
the coupling between the lipid bilayer and the spectrin filaments network
provides the ME with viscoelastic  properties. In addition to its bending
modulus $K\approx 10^{-19}-10^{-20}$  J, the RBC ME acquires two
dimensional (2D) static shear modulus revealed by micropipette suction [5] or
by other types of  static deformation response. Measurements of thermally
activated shape fluctuations [8] refine the picture and give the access to the
dynamic properties of RBC. The basic theoretical model [9] explaining both
statics and   dynamics of RBC considered the ME as practically
independent on the CS influence, but with strongly variable 2D shear
modulus. Namely, thermally excited surface undulations of RBC (the cell
flickering) [8] was  described as a motion of  a fluid ME with $\mu_s\approx0$.
Thus the free ME model [9] has established the opinion that RBC
combine paradoxally liquid-like and solid-like mechanical properties, with
solid-like behavior being manifested only during large-scale shape changes [9-10].
However, the modern microrheological technique  [e.g. 11-12] has shown
that mechanical behavior of RBC satisfies Hook's law in a broad region of
deformation [13].  This result is incompatible with both deformation
scale separation principle and non-linear shear modulus behavior
resulting from the model [9].

As it will be shown later apparent softness of RBC obtained from
the fit of thermal fluctuation spectra by the free ME model is
related to the simplified form of its elastic energy. In [9] it is
reduced to the ME bending and shear terms only. In recent
theoretical works [14-17] it was proposed to take into account the
influence of viscoelastic CS properties on the coarse-grained
mechanics of the composite ME-CS system. Energy of surface tension
and especially the term responsible for the ME confinement were
introduced in the system's Hamiltonian [16-17]. Resulting model
clarified the effective bending modulus behavior during normal
fluctuations. Besides, it demonstrated the physical factors which
limit normal fluctuations of the ME. By contrast, model [16-17]
could not explain the main RBC paradox : the apparent change of its
shear properties from the solid-like to the liquid-like ones. In
approach [16-17] the CS was treated as an infinitely rigid shell
(i.e. with an infinite shear modulus) making impossible the
calculation of the effective shear modulus for the ME-CS structure.
Infinite rigidity of the CS prevents any tangent fluctuation in the
model, and thus does not allow to compare its predictions with the
results of the free ME model [9]. Note that the conclusion about
liquid-like behavior of RBC was obtained in [9] from the
calculation of the ratio $<X_n^2>/< X_t^2>$ of mean-square
amplitudes of normal and tangent fluctuations (see Fig. 1).
To solve the problem authors of Ref. [18] proposed
to consider the dependence of $\mu_s$ value on the wave vector. In
their model $\mu_s$ modulus tends to zero only for the fluctuations
with wavelengths smaller than the mesh size of the spectrin network.
Nevertheless, the $<X_n^2> /< X_t^2>$ ratio has not been calculated.

In the present work we propose more realistic continuum model of the
composite ME-CS system which brings a solution to the paradoxal
problem of the RBC apparent softness  in thermal fluctuation measurements. We
show that the 2D shear modulus of the RBC obtained from the fluctuation data
and that  measured in static deformation experiments have the same order  of
magnitude $\mu _s\approx 6\times 10^{-6}$ N/m.
For that purpose, in addition to the ME confinement, we take into
account the contribution of normal and tangent fluctuation modes of a
finite-thickness CS to the surface free energy of the ME-CS
system. These modes, determined in the simplest case of incompressible
CS by its bulk shear modulus  $\mu$ only, are rapidly attenuated in the
cell interior. Normal CS fluctuations modify the ME bending properties
and tangent fluctuations supply the composite system
with a finite 2D shear modulus $\mu_s$.  Mean-square
amplitudes of normal and tangent fluctuations of the system are calculated in the framework of the
proposed model. They  are perfectly consistent with the RBC flicker
spectroscopy data [8,19] but their ratio $<X_n^2>/<X_t^2>$  is 2-3 orders of magnitude
smaller than that attributed in [9] to the solid-like ME.
  This result shows that the RBC fluctuational
softness is apparent and its shear properties in both static and dynamic
experiments are the same.

The material constants for these estimations were obtained by a
double-step numerical fit of the effective bending rigidity
dependence on the wave vector $q$ [19]. The same type of
experimental data is frequently used in the theoretical estimations
of the RBC mechanical properties [16,17,20]. The data being
obtained using classical Fourier transform, the models are usually
formulated in the framework of plane geometry. We also follow this
approach but limit the region of its applicability. A model which
does not take into account the curvature of the system is valid
[14-17] only for a treatment of fluctuations with the wavelengths
$\lambda<<R$, where $R\approx$ 4 $\mu m$ is the effective RBC
radius. The description of the ME-CS system motion is performed
then in terms of continuum mechanics. This approximation is favored
by the fact that the attachment points between the spectrin network
and the bilayer are rather rare and their area is negligible (Fig. 1).
Besides the actin nodes there exist additional ME-CS connections
through ankryn complexes distributed randomly. Together with
defects and dynamic structural rearrangements in the spectrin
network they contribute to the entropic elasticity of the RBC CS
and justify the continuum description. Finally, we consider the
fluctuation wavelength region for which the influence of active
processes is reduced to the increase in the effective temperature
[21-22].

The free energy of normal thermal fluctuations of a
ME-CS system contains the fluctuation energy of
a flat ME and the CS contribution
$W$:
\begin{equation}
F= \int dS \left(\frac{1}{2}K(\Delta u^m)^2+\frac{1}{2}\sigma(\nabla u^m)^2 +
\frac{1}{2}\gamma(u^m)^2 + W({\bf u}^c)\right).
\end{equation}
Here  $u^m$ is a normal displacement of the ME, $K$, $\sigma$ and $\gamma$ are
the coefficients of bending, surface tension and pinning of the ME,
respectively; ${\bf u}^c$ is  a displacement field  of the CS.

The ME fluctuation energy (first three terms in Eq. (1)) is
limited to the sum of bending, surface tension and
ME confinement energy contributions. In a plane geometry adopted in the
present work bending fluctuations are independent on stretching
and the energy due to the bending-stretching coupling
can be omitted, though in a general system with curvature it
cannot be neglected [15,23,24].
The CS is considered as an incompressible  viscoelastic plate (Fig. 1) with
the thickness $h\approx0.03$ $\mu m$ and with a finite 3D shear
modulus $\mu$.
The last term in (1) expresses  the
surface density of the CS fluctuation energy:
\begin{equation}
W= \int_0^h dz \mu\epsilon_{ij}\epsilon_{ij}^*,
\end{equation}
where $\epsilon_{ij}=\frac{1}{2}(\frac{\partial u_i^c}{\partial x_j}+\frac{\partial u_j^c}{\partial
x_i})$ is the CS strain tensor, $u_i^c=u_x^c,u_y^c,u_z^c$
and $x_i=x,y,z$; and $z$ is directed perpendicular to the
ME-CS interface. Due to the small Reynolds number
of the system the motion of both ME and CS is
overdamped. Thus the state of CS is determined from the
motion equation of an incompressible medium with
neglected inertia terms:
\begin{equation}
\nabla p= \mu \Delta {\bf u}^c +\eta \Delta \dot{{\bf u}^c}
\end{equation}
with $p$ standing for pressure, $\eta$ for the CS viscosity and
$\dot{{\bf u}^c}$ for the CS velocity field. In a more general case
of a compressible isotropic medium the energy (2) would have two
terms related to two independent elastic Lame coefficients
$\lambda$ and $\mu$. The equation of motion of a
\textit{compressible} isotropic medium would contain the term
$\lambda\text{ } grad\text{ } div\text{ }{\bf u}$, where ${\bf u}$
is a displacement field. This term stands for the density of forces
of isotropic deformation. In the \textit{incompressible} medium
$div\text{ }{\bf u}=0$ and $\lambda$ tends to infinity. As it is
shown in [25] the limit value of the term $\lambda\text{ }
grad\text{ } div\text{ }{\bf u}$ in the incompressible medium is
equal to -$\nabla p$. This leads to the motion equation of the
incompressible medium in the form given by Eq. (3).

Thermally excited CS fluctuation modes are rapidly
attenuated in the direction perpendicular to the ME plane.
The solution of Eq. (2) compatible with bending fluctuations of the
ME and satisfying $div\text{ }{\bf u}^c=0$ condition describes
the mode of this type:
\begin{eqnarray}
u_x^c&=&-i U_q q z\text{ }exp(-qz+iqx);\text{ }u_z^c=U_q (1+qz) exp(-qz+iqx);\text{ }u_y^c=0;
\\
p&=&-2iq U_q(\eta\omega+i\mu) exp(-qz+iqx);
\\
\dot{{\bf u}^c}&=&-i\omega {\bf u}^c.
\end{eqnarray}
The wave vector of fluctuations in (4-6) is chosen along the $x$
axis, $U_q$ expresses the amplitude of normal displacements at the
ME-CS interface $z=0$ (see Fig. 2). All tangent displacements
in the mode of this type vanish at the interface.

In the present work we consider the physical situation, where at the interface the
CS velocities $\dot{{\bf u}^c}$ and displacements
${\bf u}^c$ coincide with those of the ME:
$\dot{{\bf u}^c}|_{z=0}=\partial{\bf u}^m/\partial t$ and
${\bf u}^c|_{z=0}={\bf u}^m$. In a more complicated
case of ME-CS mutual gliding,
 only the normal components of both velocity and
displacement fields are preserved at the
interface and a viscous drag influences tangent
fluctuation modes. However, in the plane geometry approximation
this does not affect the normal fluctuation modes (which are of
primary interest in the RBC model) and maintains the relation
$u^m=U$, where $u^m$ and $U$ are the normal displacement of ME and
CS, respectively.

Finally, the contribution $W_q$ of the CS to surface
fluctuation energy (1) is obtained by substitution of bending
displacement field (4) into elastic energy (2) and following
integration over the CS thickness:
\begin{equation}
W_q= \mu q|U_q|^2\left( 1-exp(-2qh)(2q^2h^2+2qh+1)\right).
\end{equation}

After development of  free energy functional (1) into Fourier
series we get the total energy of normal ME-CS
fluctuations in the form:
\begin{equation}
E=S\sum _{\bf q}E(q)|U_q|^2
\end{equation}
suitable for analysis of the RBC fluctuation spectrum [8,19].
Function $E(q)$ is expressed as
\begin{equation}
E(q)=Kq^4+\sigma q^2+2W_q/|U_q|^2+\gamma.
\end{equation}
The sum in Eq. (8) runs over all possible {\bf q} in a square plate
with the side equal to $\pi R$; the interface area $S$ is taken to
be $(\pi R)^2$. Note that in the limit  $qh\rightarrow 0$ CS
fluctuations simply renormalize the ME bending rigidity: $W_q\approx
4\mu h^3|U_q|^2q^4$/3; and in the limit $qh\rightarrow \infty$
CS fluctuation wave transforms into the wave of normal fluctuations
on the surface of a semi-infinite medium: $W_q\approx \mu
q|U_q|^2$. Due to a strong exponential attenuation of solution
(4-6) with the depth $z$ the last approximation is suitable in a
wider region $q \geq h$.

The authors of [19] give the fluctuation spectra of RBC as the
dependence of $-log<|U_q|^2>q^4$ on the wave vector $q$ in the
wavelength region from 0.25 $\mu m$ to 2 $\mu m$. They stress also
that the region of fidelity of their results does not exceed 0.5
$\mu m \leq \lambda \leq$ 1 $\mu m$ limits. Long-wavelength limit
is determined by the validity of the plane geometry approximation
(used in [19] for the data processing). Lack of fidelity in the
short-wavelength region is related to the experimental technique
limitations. Therefore, we performed the fit of the RBC fluctuation
spectrum in the fidelity region only. Using equipartion
$E(q)\propto <|U_q|^2>^{-1}$ one can fit the expression $q^4/E(q)$
proportional to the experimental value $q^4<|U_q|^2>$, and usually
related to the effective bending rigidity [16,17,20]. Double-step
least-square fit (Fig. 3) of relative values of $q^4<|U_q|^2>$
leads first to the expressions of the RBC material constants in
function of bending modulus $K$. The inset in Fig. 3 shows the
relative contributions of the bending, surface tension, CS
fluctuation and confinement energies to the total mode energy
$E(q)$ for one of the three cells studied in [19]. Similar relation
between different contributions exists for two other cells. Then,
taking into account the average value $K=3.4\times10^{-20}$ J
estimated in [19] we obtain the absolute values of the constants:
$K^{(1)}\approx4.9\times10^{-20}$ J;
$\sigma^{(1)}\approx-5.9\times10^{-6}$ N/m;
$\gamma^{(1)}\approx3.1\times10^{8}$ N/m$^3$; $\mu^{(1)}\approx148$
N/m$^2$; $K^{(2)}\approx2.2\times10^{-20}$ J;
$\sigma^{(2)}\approx-2.3\times10^{-6}$ N/m;
$\gamma^{(2)}\approx1.8\times10^{8}$ N/m$^3$; $\mu^{(2)}\approx105$
N/m$^2$. The fit of experimental data [19] for the third cell (not
shown in Fig. 3) results in $\mu^{(3)}\approx165$ N/m$^2$. The
found value of the CS 3D shear modulus $\mu$ is in a good agreement
with an effective spring constant of spectrin filament
$\approx4\times10^{-6}$ N/m in 30 nm thick CS. The effective value
of the 2D shear modulus $\mu_s$ can be estimated in the simplest
way as $\mu_s\approx\mu h=(3.2\div 5)\times10^{-6} N/m$.

Let us now show that to explain the RBC behavior in thermal
fluctuation experiments we do not need to suppose vanishing of its
effective 2D shear modulus ($\mu_s\approx0$). To do this we
reexamine two main arguments of the free ME model [9] which
have lead to the conclusion about negligible $\mu_s$ value and,
then reanalyze the same experimental data in the framework of the
composite ME-CS model developed in the present
work.

The arguments developed in [9] can be resumed qualitatively in the
following way: 1)  The ratio of mean-square amplitudes of normal and tangent
fluctuations $<X_n^2>/<X_t^2>$ in the model [9] has the
order of magnitude $\sim\mu_sR^2/K$.
2) Thermal fluctuation experiments show that the
RBC thickness fluctuation profile has the peak near the rim of the
cell [8]. Due to the RBC shape the main contribution to the thickness
fluctuations near the rim comes from tangent fluctuations (see
Fig. 1). Thus, in a system with $\mu_s\approx0$ and, consequently,
with strong tangent fluctuations, the peak should find itself near the
rim. With the value of $\mu_s\approx6\times10^{-6}$ N/m known from
static experiments the  model [9] gives the $<X_n^2> /< X_t^2>$
ratio
2-3 orders of magnitude greater than the experimental
one. The discrepancy
is then attributed to the high value of $\mu_s$ and to preserve
the model the effective 2D shear modulus is considered to be
vanishing.

Estimation of the $<X_n^2> /< X_t^2>$ ratio in the framework of
the present model gives however quite different result which is
consistent with the static $\mu_s$ value.  The main difference
with respect to the model [9] is determined by
the contribution to total energy (8) from the CS,
both from its fluctuations and from the ME confinement.
Mean-square amplitudes of normal and tangent fluctuations of the
composite ME-CS system can be, in principle,
calculated using equipartition theorem. To avoid however, a
cumbersome discussion of different mode polarizations for tangent
fluctuations we prefer to illustrate the results in the way
proposed in [26] and based on the fluctuation-dissipation
theorem.
Mean-square amplitudes of normal and tangent fluctuations are then
determined by static normal and tangent response functions,
respectively [26,27]:
\begin{equation}
<X_i(0)X_i(0)>=K_BT\alpha_i(0),
\end{equation}
where response $\alpha_i(\omega)$ defines
amplitude $A_i(\omega)$ of ME particle motion under
periodic external force $F_i=F_i^0exp(-i\omega t)$ application:
$\alpha_i=A_i/F_i^0$. The values of responses at zero
frequency, and consequently, the
amplitude of mean-square fluctuations are independent on the
system dissipative properties. The responses of a flat
ME have been discussed in detail in [14,15]. Using the same
formalism we obtain normal response $\alpha_n(0)$ of the
composite ME-CS system in the form:
\begin{equation}
\alpha_n(0)=\int_{q_{min}}^{q_{max}}
\frac{qdq}{2\pi E(q)},
\end{equation}
where $E(q)$ is defined by Eq. (9).
The minimal wave vector $q_{min}$ of fluctuation which contributes
to response (11) is determined by the RBC finite size:
$q_{min}=1/R$.  Upper cut-off $q_{max}$ is defined [27] by the fact that
the number of fluctuation modes in the system is finite. In the
lipid ME it is determined by the total number $N$ of lipid
molecules: $q_{max}=\sqrt{N}q_{min}$. In such a case the number of
normal fluctuation modes is equal to $N$.

The expression for the tangent response $\alpha_t(0)$ is
obtained along the same line. In the general case the function  $\alpha_t(\omega)$ can be
presented as
$\alpha_t(\omega)=\alpha_t^{sh}(\omega)+\alpha_t^{st}(\omega)$,
where $\alpha_t^{sh}(\omega)$ and $\alpha_t^{st}(\omega)$ are
contributions of the shear and stretching fluctuation modes,
respectively [14].  However, the lipid ME is commonly considered
as incompressible. In such a case
$\alpha_t^{st}(\omega)=0$ and
$\alpha_t(\omega)=\alpha_t^{sh}(\omega)$. Note, that shear fluctuations in
the proposed model depend mainly on the CS properties.
To obtain the energy of shear modes we use the
solution of Eq. (3) compatible with shear fluctuations of the
ME:
\begin{equation}
u_x^c=0;\text{ }u_y^c=U_q exp(-qz+iqx);\text{ }u_z^c=0;\text{
}p=0; \dot{{\bf u}^c}=-i\omega {\bf u}^c.
\end{equation}
Similar to mode (4-6), thermally excited shear
CS fluctuations are rapidly attenuated in the
$z$-direction (see Fig. 2). Substitution of displacement field (12) into
elastic energy (2) and integration over the CS thickness
results in the shear fluctuation energy
\begin{equation}
E^{sh}=S\sum _{\bf q}\mu q|U_q|^2\left(1-exp(-2qh)\right).
\end{equation}
Corresponding shear contribution to the tangent
response of the composite system is then expressed as:
\begin{equation}
\alpha_t^{sh}(0)=\int_{q_{min}}^{q_{max}}
\frac{dq}{4\pi \mu(1-exp(-2qh))}.
\end{equation}

To calculate responses (11) and
(14) we use the RBC material constants $K$, $\sigma$, $\gamma$ and
$\mu$ fitted above and take $N
\approx4.4\times10^{7}$ molecules [28]. The normal response $\alpha_n(0)$ is
estimated to be $\sim(6\div9)\times10^{4}$ m/N. This value is 2-3
orders of magnitude smaller than the response function of the free
ME model with the same bending rigidity. For the tangent response
function $\alpha_t(0)$ we obtain $\alpha_t(0)$
$\sim(8\div11)\times10^{5}$ m/N. Thus, final $<X_n^2>/< X_t^2>$
ratio is also 2-3 orders of magnitude smaller than that in the free
ME model and has no contradiction with the thickness fluctuation
profile experiment. Let us stress, that the value obtained in the
approximation of an incompressible ME-CS system is an upper limit
for the $<X_n^2>/< X_t^2>$ ratio. The system compressibility
increases the $< X_t^2>$ value and preserves the $<X_n^2>$ one. It means
that for a compressible system our approach works even better.

Knowledge of responses (11) and (14) allows us to make a little
better estimation of $\mu_s$ modulus than that given by the relation
$\mu_s\approx\mu h$. The following estimation is more appropriate for
the static microrheological experiments using the
probe particles technique (see, for example [13]). In such experiments
the cell is usually replaced for data processing by a 2D
solid-like ME. The static shear properties of the
both systems are roughly equivalent provided
$\alpha_t^{sh}(0)=\beta_t^{sh}(0)$, where $\beta_t^{sh}(0)$ denotes
the shear contributions to the tangent response of the 2D
solid-like ME. The response
$\beta_t^{sh}(\omega)$ of the 2D ME in connection
with the PP microrheology was studied in [14,15]. Its value
\begin {equation}
\beta_t^{sh}(0)=\int_{q_{min}}^{q_{PP}}\frac{dq}{4\pi q\mu_s}
\end {equation} depends
on the wave-vector cut-off $q_{PP}=2\pi/R_{PP}$ [14,15], where
$R_{PP}$ is the radius of the contact area between the RBC and the
PP (i.e. silica bead). The equality between $\alpha_t^{sh}(0)$ (see Eq.
(14)) and $\beta_t^{sh}(0)$ results in the following effective 2D
shear modulus $\mu_s$:
\begin{equation}
\mu_s\approx\mu ln\left(\frac{2\pi R}{R_{pp}}\right)/\left(\frac{2\pi}{R_{pp}}+
\frac{1}{2h}ln\frac{(1-exp(-4\pi h/R_{PP}))R}{2h}\right).
\end{equation}
Taking $R_{PP}\approx R/20$ we obtain that the CS with
the bulk shear modulus $\mu\approx 105\div 165$ Pa and $h\approx 30$
nm thickness induces in the composite ME-CS system
the 2D shear modulus $\mu_s\approx (5.1\div 8.1)\times10^{-6}$
N/m. This value is very close to that obtained in static
deformation experiments [5-7].

In addition to the resolution of the RBC apparent softness
paradox, the composite ME-CS model has a striking
feature: negative $\sigma$ value obtained from the fit of
fluctuation spectra [19].
Negative osmotic pressure difference $\Delta P=2\sigma/R
\approx1\div1.5$ Pa insures a good
contact between the ME and the CS. It makes the mechanical
interaction between the parts of the composite system much less
dependent on the attachment points. In such a case the CS induces
(through the controlled tension mechanism) additional ME
fluctuations which can even result in an elastic instability of the
cell [29]. A high negative surface tension would break the symmetry
from the normal biconcave RBC to the echinocyte shape. On the other
hand, a high positive surface tension corresponding to a
sufficiently big positive osmotic pressure difference would make a
normal cell unstable with respect to the spherical shape. These
points attract considerable attention in the cell biology field
[30,31].

More detailed physical discussion of the surface tension could be
done in the frame of a more rigorous model taking into account both
finite shear modulus $\mu_s$ and the real RBC shape. In such a
model surface tension becomes a tensor characteristic $\sigma_{ij}$ of
the ME that satisfies the Laplace equilibrium equation:
$\sigma_{\theta \theta }/R_{\theta}+\sigma_{\phi \phi
}/R_{\phi}=\Delta P$. Here $R_{\theta}$ and $R_{\phi}$ are the RBC
curvature radii along $\theta$ and $\phi$ spherical coordinates,
respectively. Due to the rotational symmetry of the non-pathological RBC
shape $\sigma_{\theta \phi }=0$.
The values of the $\sigma_{\theta \theta }$ and $\sigma_{\phi \phi
}$ components are dependent on the $\theta$ coordinate of the ME
point. Thus, change in signs of principal curvatures $1/R_{\theta}$
and $1/R_{\phi}$ can alternate compressed and stretched regions at
the cell surface. If $\Delta P$ is negative (as it is obtained in
the present model) then the ME region near the rim is
compressed: positive curvatures near the rim lead to $\sigma_{\theta \theta}<0$ and
$\sigma_{\phi \phi}<0$. This fact favors normal fluctuations and
increases their contribution to $<\delta d^2>$ (see Fig. 1). Near
the cell center the RBC shape is locally concave and both
curvatures are negative. The ME is stretched in this region
($\sigma_{\theta \theta}>0$ and $\sigma_{\phi \phi}>0$) and its
normal fluctuations are reduced. Consequently,  the amplitude of normal
fluctuations increases with the distance from the cell center. Remarkably,
it puts negative osmotic pressure value in a good agreement with
the location of the thickness fluctuation profile peak.

In conclusion, we showed that the RBC paradox, consisting in
solid-like elastic behavior in static suction and liquid-like
behavior in thermal fluctuations experiments is apparent. The
influence of the finite-thickness CS which confines the
ME and contributes to the fluctuation energy of the composite
ME-CS system reduces considerably normal
fluctuations of the RBC. Contrary to the predictions of the free
ME model, the amplitude of normal fluctuations in the composite
system is smaller than the amplitude of tangent fluctuations.
Resulting effective 2D shear modulus of the RBC
shows in thermal fluctuation spectra the value typical for the
solid-like elastic behavior.

The authors thank G. Mennessier for helpful discussions. S.B.R. is
grateful to the Biannual Program of the French Minestere de
l'Education Nationale, de l'Enseignement Superieur et de la
Recherche for financial support.

\begin{center}
{\Large Figure captions}
\end{center}

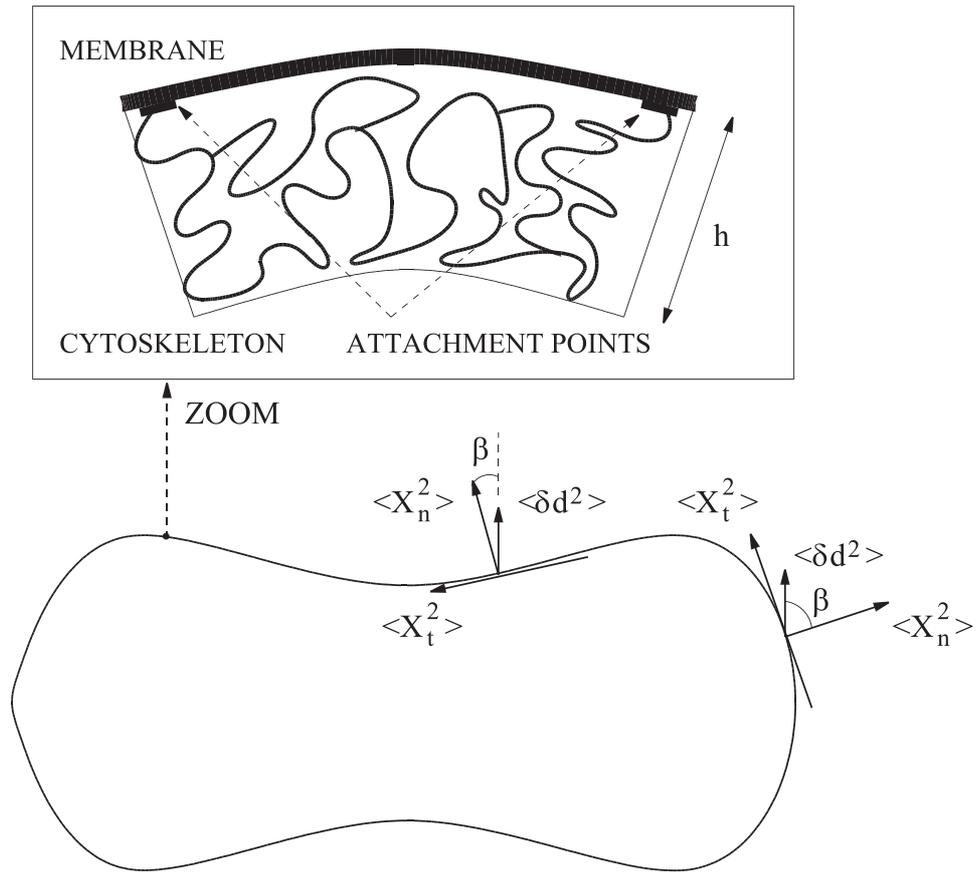
Fig. 1. Equilibrium shape and fluctuations of the RBC. Mean-square
amplitude of the RBC thickness fluctuations $<\delta d^2>$ depends
on the mean-square amplitudes of normal $<X_n^2>$ and tangent  $<X_t^2>$
fluctuations and on the cell surface point: $<\delta
d^2>=<X_n^2>cos^2\beta+<X_t^2>sin^2\beta$;  here $\beta$ is the angle
between the vertical direction and the RBC surface normal.  Near the rim of the
cell the main contribution to $<\delta d^2>$ comes from tangent
fluctuations. Inset: Schematic representation of the RBC membrane
and finite-thickness cytoskeleton.

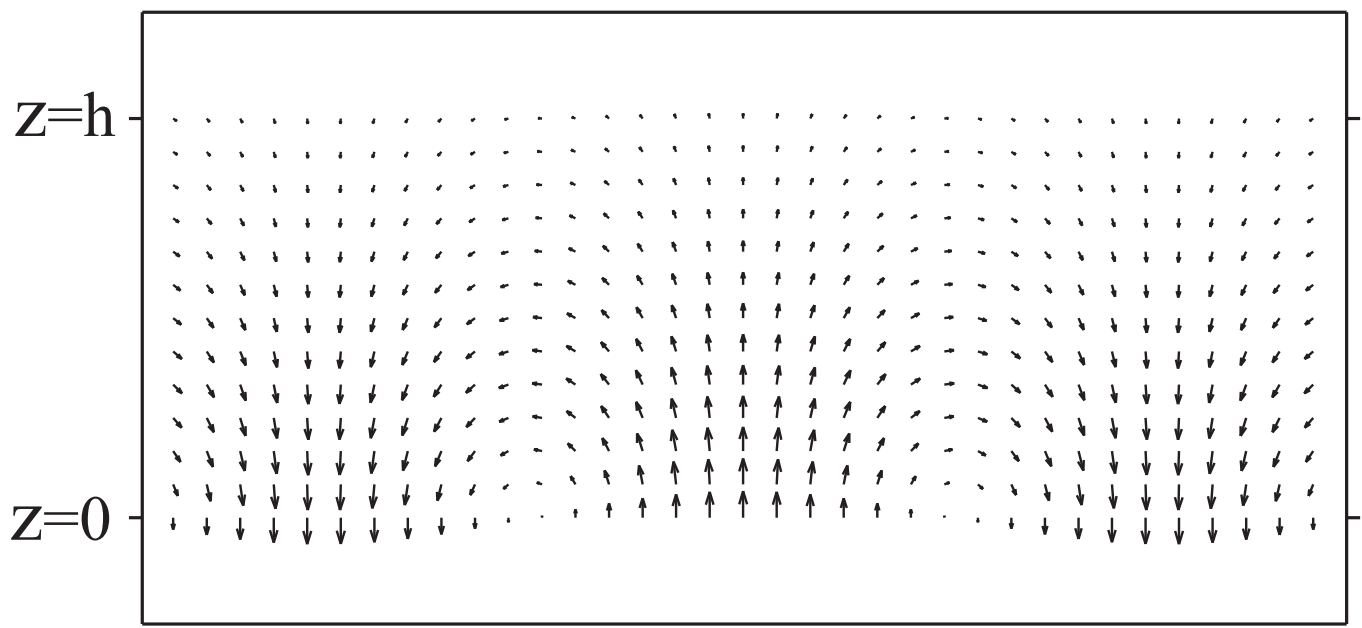
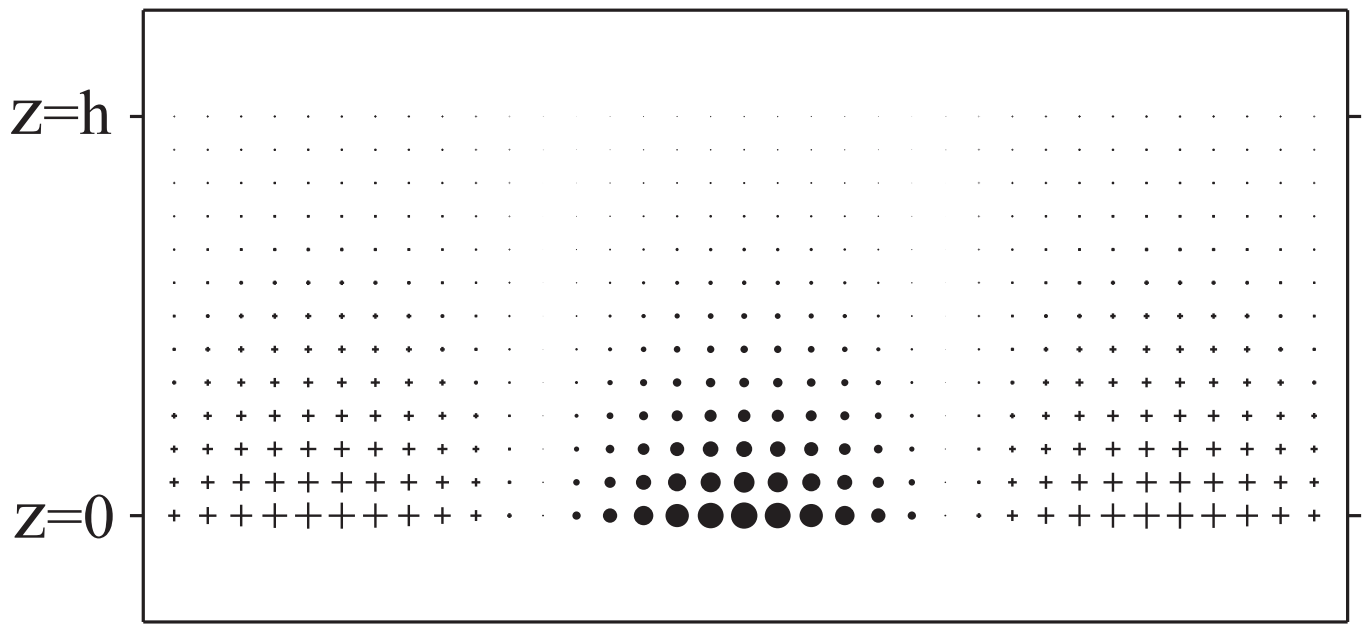
Fig. 2. Displacement fields in the cytoskeleton. 2D sections by the
plane containing $z$ and $x$ axes are shown. The three different
modes presented are characterized by the same wave vector ${\bf q}$. The
membrane is located in $(x,y)$ plane at $z=0$ level. According to
the boundary condition (see in the text), the displacement of the
membrane surface is equal to that of the cytoskeleton at $z=0$ level,
therefore the membrane is not shown. (a) Bending mode.
(b) Shear mode. The displacement field is
perpendicular to the plane of the figure. Two opposite directions
of the displacement field are shown by crosses and full circles.
Their size is proportional to the displacement value.

Fig. 3. Wavelength dependence of $q^4/E(q)$ ($E(q)$ is given by Eq.
(9)) related to the RBC effective bending rigidity in the flicker
spectroscopy experiments [19]. Fits (solid lines) of the data for
two different cells (rhombuses and squares) are presented. Inset:
Calculated wavelength dependence of the terms in the normal
fluctuation mode energy $E(q$): bending contribution $E_1=Kq^4$,
surface tension contribution $E_2=\sigma q^2$, CS fluctuations
contribution $E_3=2W_q/|U_q|^2$ and confinement $E_4=\gamma$.
Though the surface tension term is negative, the globally
positive mode energy $E(q)$ insures stability of the system with
respect to fluctuations.

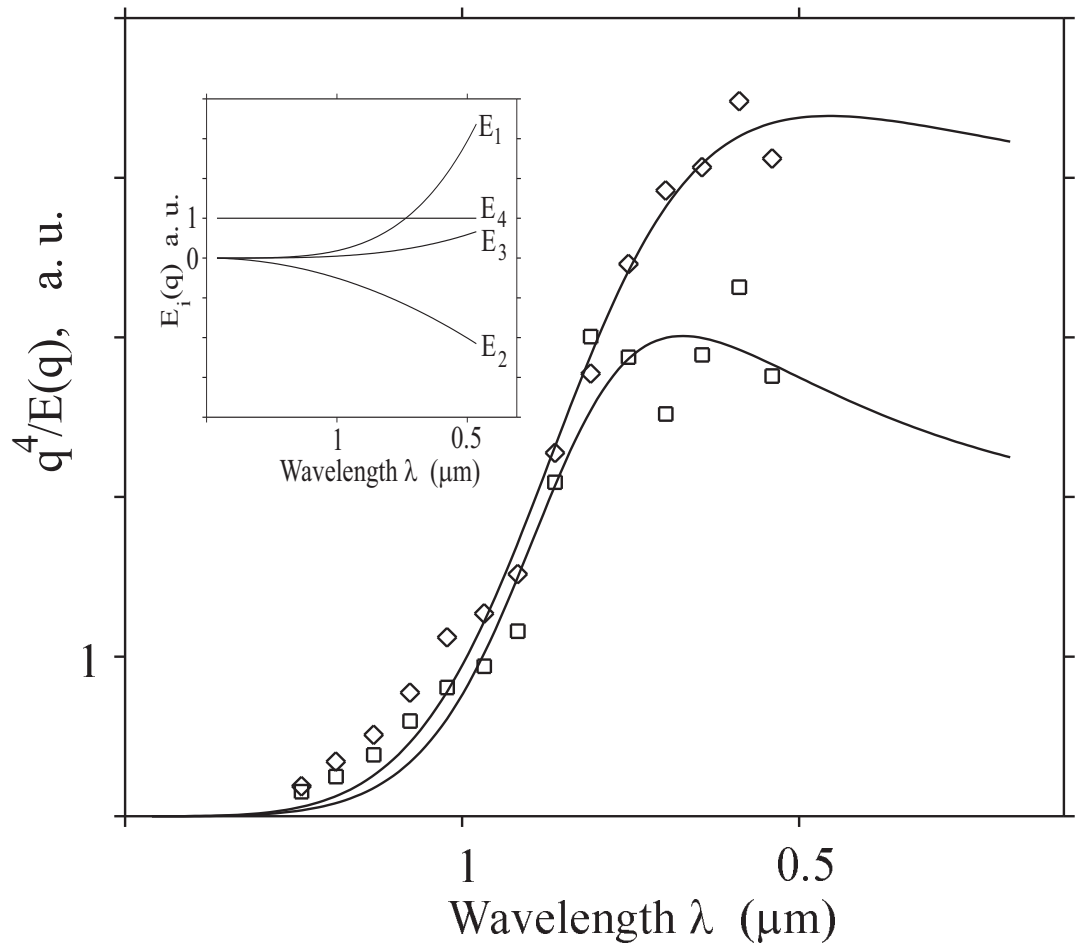
\end{document}